# FIELD STARS AND CLUSTERS OF THE GALACTIC BULGE: IMPLICATIONS FOR GALAXY FORMATION [1]

Dante Minniti[2]

*European Southern Observatory, D–85748 Garching b. München, Germany*

*and Steward Observatory, University of Arizona, Tucson AZ 85721*

## ABSTRACT

The results of a kinematic study of the Galactic bulge based on spectra of red giants in fields at projected distances of 1.4–1.8 kpc from the Galactic center are presented. There is a marked trend of kinematics with metallicity, in the sense that the more metal–poor population has higher velocity dispersion and lower rotation velocity than the metal–rich population. The K giants more metal–poor than [Fe/H] $= -1$ have halo–like kinematics, with no significant rotation and $\sigma \sim 120$ km s$^{-1}$ independent of Galactocentric distance. The velocity dispersion of the giants with [Fe/H] $\geq -1$ decreases with increasing Galactocentric distance, and this population is rotating with $V \sim 0.9$ km s$^{-1}$ degree$^{-1}$.

The present observations, together with the observed metallicity gradient imply bulge formation through dissipational collapse. In such a picture, low angular momentum gas lost from the formation of the halo was deposited in the bulge. The bulge would therefore be younger than the halo. Observations of the Galactic globular cluster system are consistent with this picture, if we associate the metal–rich globular clusters within 3 kpc of the Galactic center with the bulge rather than with the thick disk or halo. Data on the RR Lyraes in Baade's window are also consistent with this picture if these stars are considered part of the inner halo rather than the bulge.

*Subject headings:* Galaxy: kinematics and dynamics — Galaxy: Structure — Galaxy: formation

---

## 1. Introduction

The process of formation of our Galaxy (or indeed any other spiral) is poorly understood. Searle & Zinn (1978) proposed an inhomogeneous halo formation in which small protogalactic fragments coalesced over an extended period of time. This picture is more widely accepted today than the alternative model of halo formation via a rapid dissipational collapse, originally proposed by Eggen, Lynden–Bell & Sandage (1962, hereafter ELS), and explored theoretically by Larson (1976). Zinn has recently (1993) made a hybrid model, in which some of the halo clusters formed in a dissipational collapse, and others, especially at large Galactocentric distances, came from coalesced fragments. In the past the formation of the Galactic bulge has not received as much attention as the formation of the halo, perhaps because it has been assumed that they are parts of a single component (the spheroid), or because of crowding and reddening. Our Galaxy offers us the best hope of reliably discriminating between different components (bulge, disk and halo), and of disentangling the chronology and mechanism of formation. For more distant galaxies, one has had to rely on model decompositions of the integrated properties.

In this paper we summarize the main results from Minniti (1993). The goals of this work are to determine how the Milky Way bulge formed, and to determine how the bulge fits into the stellar populations of the Galaxy. The approach taken is to measure kinematics and abundances for large numbers of bulge giants in selected off–axis fields, and to study the clusters that might be associated with the bulge.

Based on those data, we defend the idea that the Galactic bulge was formed by dissipational collapse. We also argue (following Carney 1990 and Wyse & Gilmore 1992) that the bulge formed from the gas left over by the formation of the halo.

## 2. Dependence of Kinematics on Metallicity

Previous reports of the dependence of kinematics on metallicity for bulge giants are given by Rich (1990), Minniti et al. (1991), Spaenhauer, Jones & Whitford (1992), Harding & Morrison (1993), and Zhao et al. (1995). While the statistics are poor and all of these authors do not agree in all points, a general conclusion is that there is a trend of kinematics with metallicity, in the sense that the more metal–poor stars have larger velocity dispersions than the more metal–rich ones.

We have completed a study of 3 bulge fields, obtaining medium resolution spectra for ∼700 giants. The observations were carried out in several runs in the Steward Observatory 2.3 m telescope with the MX multifiber spectrograph, the CTIO 4 m telescope with the ARGUS multifiber spectrograph, the Multiple Mirror Telescope with the Red Channel spectrograph, and the 100 inch telescope at Las Campanas Observatory (LCO) with the Schechter spectrograph. The observations at LCO were kindly taken by Drs. A. McWilliam and M. Rich. The data and analysis for two of the fields at $l, b = (8°, 7°)$ and $(12°, 3°)$ are given by Minniti et al. (1992), Minniti et al. (1995a), and Minniti (1995a). The candidates in these fields were selected on the basis of $B_j RI$ photometry (see Minniti et al. 1992). The $B_j - R$ vs $R - I$ color–color diagram was used to discriminate possible foreground unreddened dwarfs. The other field is centered on the globular cluster M22, at $l, b = (10°, -7.6°)$, with 100 stars observed. The data and results for this field are given by Minniti et al. (1995b). We chose here a suitable sample of bulge giants based on the location of stars in the color–magnitude diagram of Cudworth (1986). All the stars with (B–V) $\geq 0.8$, V$\leq$14, and cluster membership probability smaller than 99% were included in our candidate list. The proper motion information given by Cudworth (1986) was used as an additional check to discriminate possible unreddened nearby dwarfs.

More than 90% of the stars observed in all the fields are K giants, the rest being M giants. For the K giants we derive abundances by measuring the strength of spectral indices as described by Minniti (1995c), calibrating them against a grid of giants observed in clusters of well known abundances. We also measure radial velocities with $1\sigma$ errors of $\sim$ 10 km s$^{-1}$ for all K and M stars.

These data show a strong dependence of the kinematics on metal abundance. Figure 1 illustrates this effect for the field F588 at $l, b = (8°, 7°)$, which has the smallest disk contamination and the largest number of stars analyzed. The abcissa corresponds to the globular cluster metallicity scale of Armandroff (1989).

Even though there is no dramatic break in these diagrams that would suggest a discrete change between populations, it turns out to be profitable to examine the kinematics of stars in two groups: stars more metal–rich than [Fe/H] $= -1.0$, and metal–poor stars with [Fe/H] $\leq -1.0$.

We have rated the significance of the kinematic differences between these two groups by drawing random subsamples of stars from our entire observed catalogue. The set of observed velocities and observed abundances were paired at random, and the distribution of kinematics as a function of metallicity was derived. The metal–poor group in field F588 has a smaller mean velocity by $2\sigma$, and larger velocity dispersion by $2.5\sigma$, than the metal–rich group. Differences as large as these were seen in only 24 out of $10^5$ trials. Thus, the dependence of kinematics on metallicity is significant at a high confidence level.

The available proper motion data for bulge fields show trends of tangential velocity dispersion with metallicity similar to the radial velocity components (Minniti et al. 1991, Spaenhauer et al. 1992). It is found that the proper motion components of the velocity dispersions for metal–poor stars are larger than those of the metal–rich stars in the bulge, both in Baade's window (Spaenhauer et al. 1992), and the M22 field (Minniti et al. 1991), at 0.5 and 1.7 kpc from the Galactic center, respectively.

The mean velocities and velocity dispersions measured for metal–poor and metal–rich K giants are listed in Table 1 for the different fields. The dependence of kinematics on Galactocentric distance is shown for both metal–poor and metal–rich stars in Figure 2. In this figure we have added data for other fields from a variety of sources. The innermost point comes from the data of Rich (1990), after separation of the stars according to our division into metal–poor and metal–rich components, and after scaling them to the metallicity scale of McWilliam & Rich (1994). The field at 2 kpc studied by Harding & Morrison (1993) is also included, as are the results of Morrison (1993) for a field at 5 kpc, and of Harding (1990) for a field at 3.2 kpc (there are no metallicities measured for this field, but is far out enough in the halo that we assume halo kinematics dominate here). The metal poor panels also include data for the Solar neighborhood from Norris (1987).

A remarkable observational fact is that the global halo kinematics –as measured using halo globular clusters– are similar to the local halo kinematics –as measured from the metal–poor stars in the Solar neighborhood– (e.g. Frenk & White 1980). This suggests that the halo kinematics do not change appreciably with position in the Galaxy.

Norris (1987) argued that the halo is nearly isothermal outside the Solar circle, with slow rotation. We confirm that this is true in the inner regions of the Galaxy as well. Figure 2 again shows very clearly the different kinematics of metal–rich and metal–poor giants. The metal–poor stars show negligible rotation all the way to the center. Their velocity dispersion does not increase significantly within the inner 3 kpc. Thus, the metal–poor stars in the inner Galaxy appear to form a purely dispersion–supported system, with near zero net rotation and large velocity dispersion.

Note that Zinn (1985) concludes that the mean [Fe/H] for halo clusters does not vary much with distance. Also, Carney, Latham & Laird (1989) and Norris & Ryan (1991) show that the field halo stars in the Solar neighborhood (defined by kinematics) have few representatives with [Fe/H] $\geq -1$. Furthermore, the RR Lyraes in Baade's window are metal–poor, with mean $[Fe/H] = -1$ (Walker & Terndrup 1991), and have also a large velocity dispersion, $\sigma = 130$ km s$^{-1}$ (Gratton 1987, Tyson 1991).

The velocity dispersions and rotations of the metal–poor stars observed in bulge fields have similar values to those of the nearby halo stars, to those of the halo globulars and to those of the bulge RR Lyraes. On the basis of this comparison, it is then plausible to conclude that the metal–poor stars in this study, and the RR Lyraes in Baade's window, are the extension of the halo population within 3 kpc.

Given this evidence for an extension of the metal–poor halo into the inner part of the Galaxy, the natural conclusion would be to identify the metal–rich field stars with the Galactic bulge. This is less straightforward, since we cannot discard the possibility that interlopers from the inner disk are included. Simple Galactic models, using the best available scale lengths, scale heights and normalizations, give negligible disk dwarf contamination.[2]

The issue of disk giant contamination is more problematic, as the disk density at small Galactocentric distances is poorly known. The same simple Galactic models indicate that the percentage of disk giant stars in field F588 is minimal ($\leq 5$ %), as is generally thought to be the case in Baade's window (e.g. Terndrup 1988). Further support to our estimate of negligible disk contamination in the F588 sample is recently given by Paczynski et al. (1994) and Kiraga & Paczynski (1994), who model the number density of disk stars as constant up to ~4 kpc, and vanishing beyond that distance, as if the disk were hollow in the inner few kiloparsecs. Moreover, the observed kinematics of the metal–rich stars are different from those expected for the inner disk in this field. The disk velocity dispersion in these fields should be of the order of $\sigma_D = 100$ km s$^{-1}$ (Lewis & Freeman 1990), and the disk rotation should be $V_r = 220$ km s$^{-1}$ (Burton & Gordon 1978).

The metal–rich population has substantial rotation, and we will henceforth identify these stars with the bulge, noting again that a small amount of disk contamination may be present. The bulge velocity dispersion decreases away from the Galactic center (Minniti et al. 1992). This result is confirmed by observations of large samples of M giants in fields along the minor axis (Terndrup 1993). The kinematics of the M giants observed in all of our off axis fields are similar to those of the metal–rich K giants, with lower $\sigma$ and higher rotation than the metal–poor K giants. Kent (1992) modeled the bulge as an oblate isotropic rotator, predicting mean velocities and velocity dispersions as function of position. Comparing directly these predictions with the observed bulge kinematics reveals an excellent agreement, confirming our previous results (see DeZeeuw 1993).

---

[2]The models are described by Minniti (1993), Minniti (1995a), and Minniti et al. (1995a). They consist of a halo, a bulge, a thin disk and a thick disk. The halo is spherical and has a density $\propto r^{-3.2}$. The bulge is triaxial, with density profile given by model G2 of Dwek et al. (1995). The thin disk is modeled as a double exponential, with scale–height $h_r = 3.5$ kpc, and scale–length $h_Z = 0.35$ kpc. The thick disk is also modeled as a double exponential with the same scale–length, and scale–height $H_Z = 1.0$ kpc. The normalizations at the Solar neighborhood were taken to be $N_{thin\ disk} : N_{thick\ disk} : N_{halo} = 1.0 : 0.02 : 0.002$. The bulge is normalized according to our own IR star–counts in several bulge fields, and to predict the observed fraction of metal–poor stars in F588 at 1.5 kpc.

Note that the present definitions of halo and bulge are working hypotheses. Similar definitions are employed by Harding & Morrison (1993). However, with the data available it is not possible to decide if the bulge and the halo formed as distinct components, or if there is a smooth transition between them. A causal connection between their formation is suggested below, and overlap in properties such as their ages is to be expected.

## 3. The Globular Clusters

Zinn (1990) concluded that the inner metal–rich globular clusters have different kinematics and metallicities from the bulge. Minniti (1995b) recently argued that there are two new developments that enable us to postulate the association of metal–rich globulars within 3 kpc of the Galactic center with the bulge rather than the halo or thick disk:

1) We have measured the behavior of rotation and velocity dispersion of bulge giants. These are consistent with the kinematics of metal–rich globulars published by Zinn (1985) and Armandroff (1989). Specifically, Zinn (1990) gives $\sigma = 77 \pm 14$ km/s for the disk globulars within 15 degrees (or 2.1 kpc) of the Galactic center. This is very similar to the velocity dispersion of the metal–rich (bulge) giants, which range from $\sigma = 100$ km s$^{-1}$ at 0.5 kpc to $\sigma = 65$ km s$^{-1}$ at 1.8 kpc from the Galactic center. For the metal–poor globulars in the same region, Zinn (1990) gives $\sigma = 126 \pm 20$ km/s, very similar to the values we find for the metal–poor (halo) giants in the direction of the bulge ($\sigma \approx 120$ km s$^{-1}$. At the time of Zinn (1985), the bulge star velocity dispersion was not well known.

2) The bulge metallicity scale is 0.5 dex more metal–poor than previously thought (McWilliam & Rich 1994). The mean abundances at Baade's window and F588 are [Fe/H] = $-0.25$, and $-0.50$, respectively. The metal–rich cluster population defined by Zinn (1985) and Armandroff (1989) have metallicities $-0.8 \leq [Fe/H] \leq +0.3$. Consequently, the metal–rich globulars and the bulge giants overlap in metallicity. The lack of overlap was an important part of Zinn's argument.

These two results, coupled with the fact that the metal–rich globulars are much more concentrated to the Galactic center than the metal–poor globulars (Frenk & White 1982), lead us to propose that the clusters with R $\leq$ 3 kpc and [Fe/H] $\geq -0.8$ belong primarily to the bulge population, rather than a thick disk component (Minniti 1995b).

## 4. Bulge Formation via Dissipational Collapse

Having summarized the data and the correlation between metallicity and kinematics for both field stars and clusters, it is appropriate to address what they tell us about the formation of the Galaxy and the bulge in particular.

Figure 3 shows the location of the halo and the bulge in a classical V/$\sigma$ vs $\epsilon$ diagram (Binney 1978), where V is the peak rotation, $\sigma$ is the central velocity dispersion, and $\epsilon$ is the flattening ratio. This figure suggests that somehow during the formation of the Galaxy there was a change from an extended, pressure supported halo, to a flattened, more concentrated and rapidly rotating bulge. This suggests the following formation scenario. Only a small fraction of the gas in the proto–halo formed stars (e.g. Hartwick 1976). The rest of the gas was lost from the halo, sinking deep into the potential, due to energy loss by radiation and cloud–cloud collisions, but conserving angular momentum. The gas collapsed towards the central parts of the Galaxy, due to its low angular momentum (Carney 1990, see also Wyse & Gilmore 1992). In this model the bulge would have formed after the old halo stars (i.e. after most of the metal–poor RR Lyraes and globular clusters with blue horizontal branches). This scenario resembles the model of bulge formation and chemical evolution presented by Köppen & Arimoto (1990), which assumes that the bulge is formed rapidly by infall of gas from the halo, with star formation terminated later by SN driven winds (c.f. Arimoto & Yoshii 1987). Further chemo-dynamical models by Hensler & Samland (1995) predict a very extended period of star formation in the bulge. Clearly, measurements of the age range of the bulge population are needed to decide among these alternatives.

Radial metallicity gradients and correlations between kinematics and metal abundance are expected from dissipational collapse formation (e.g. ELS, Larson 1976). This scenario of formation for the Galactic bulge is supported by the following pieces of evidence:

1) Our data show clearly a strong dependence of kinematics on metallicity (Figures 1 and 2).

2) The strong metal abundance gradient in the inner Galaxy (Terndrup 1988, Frogel et al. 1990, Minniti et al. 1995b), and in bulges of other spiral galaxies (Balcells & Peletier 1994) might be understood as superposition of a centrally concentrated metal–rich bulge on a metal–poor and more extended halo. A strong metallicity gradient is not seen in the halo component for R $\geq$ 6 kpc (Zinn 1985, Armandroff et al. 1992), or indeed at any radius once the bulge component is evaluated.

3) The globular clusters are also consistent with this scenario of bulge and halo formation. Recently, Zinn (1993) concludes that the inner "old halo" clusters (those with R

≤ 6 kpc, and blue horizontal branches) show the signature of dissipation – larger rotation velocity, smaller velocity dispersion, and increased flattening –, and that there is a smooth transition from these "old halo" to the "disk" globulars, suggesting that they are part of the same collapse. By analogy, we argue that the field halo stars and the bulge stars show the same dissipation signature, and that it would be natural to associate the globulars with these respective components. Thus, the "disk" globulars within R ≤ 3 kpc are to be associated instead with the dominant metal–rich field stellar population, the bulge component (Minniti 1995b).

A foretaste that the Hubble Space Telescope will contribute significantly in this area is given by the ages of metal–rich globular clusters recently determined by Fullton et al. (1995) and Ortolani et al. (1995). Fullton et al. measure the age of NGC 6352, a thick disk cluster ($R = 4$ kpc). Ortolani et al. (1995) measure the ages of NGC 6553 and NGC 6528, two bulge clusters with nearly Solar metallicity located in Baade's window ($R = 0.5$ kpc). These ages are smaller or equal to that of 47 Tuc, depending on the adopted $M_{HB} - [Fe/H]$ relation.

4) The kinematics and metallicities of the RR Lyrae population in Baade's window are consistent with a simple extension of the halo properties rather than with the bulge itself. Furthermore, the production of RR Lyraes by a metal–poor population like the halo outnumbers by a factor of ∼ 50 that of a metal–rich population like the bulge (Suntzeff, Kinman & Kraft 1991). Lee (1992) argued that because the RR Lyraes found in Baade's window are metal–poor, the bulge must be older than the halo. However, aside from their presence in Baade's window, there is no reason to think of them as true bulge stars, especially given the kinematics of other metal–poor stars discussed above. Note that the RR Lyraes measured by Rodgers (1977) in a field at 2 kpc have low velocity dispersion $\sigma = 62 \pm 10$ km s$^{-1}$, but they belong to the thick disk rather than the bulge (Rodgers 1989). We then prefer to associate the "bulge" RR Lyraes with the halo. In this case there is no contradiction between the old ages derived from the RR Lyraes (Lee 1992) and the younger age derived for the bulge (e.g. Terndrup 1988, Holtzmann et al. 1993). Thus, the inner halo may be older than the outer halo, but the age of the bulk of the bulge population cannot be determined from the available RR Lyrae data.

5) A currently weaker piece of evidence (awaiting more data) in favor of dissipational collapse comes from the element ratios measured in bulge giants (McWilliam & Rich 1994). These appear to differ from the values typical both of disk stars (Edvarsson et al. 1993), and of halo stars (Wheeler, Sneden & Truran 1989). The bulge K giants show higher than solar abundances of Al, Mg and Ti with respect to Fe, but solar relative abundances of Ca and Si with respect to Fe. However, note that recent measurements of the composition of

915960d1d4a47411ecegiants in the metal–rich cluster M71 show similar abundance patterns to the bulge giants (Sneden et al. 1993). The Eu, Al, Mg and Ti (and possibly O) enhancement in bulge giants suggests that enrichment of the bulge was fast, with SN type II as the major source of heavy elements (c.f. Matteucci & Brocato 1990, Köppen & Arimoto 1990).

In this context, it is also interesting to compare the Milky Way bulge with the bulges of other spirals. Dissipational collapse models suggest not only the presence of abundance gradients (Larson 1976), but also that these abundance gradients may increase with luminosity (Carlberg 1984). Balcells & Peletier (1994), based on a photometric study of dust–free spiral bulges, confirm that metallicity gradients exist in spiral bulges, and that these gradients increase with total bulge luminosity. In particular, they note that for the bulges with $M_v \leq -20$ the observations are consistent with the available dissipational collapse models. The Milky Way bulge fits within this lower luminosity category.

Low luminosity elliptical galaxies are similar to spiral bulges in many respects, and may have formed in the same way (e.g. Whitford 1978, Frogel 1988, Franx 1993). How does the Milky Way bulge compare with other "ellipticals"? Carollo, Danziger & Buson (1993) find that low mass elliptical galaxies show a correlation between abundance gradients and total mass. Based of this evidence they argue that ellipticals with $M \leq 10^{11} M_\odot$ form by dissipative collapse.

## 5. Conclusions

We conclude that the bulge was formed by a dissipative collapse, from material left over after the formation of the halo, and that the metal–rich clusters within $\sim 3$ kpc of the Galactic center are associated with the bulge rather than with the thick disk or halo.

Specific predictions of the present scenario are: 1) The RR Lyrae population in the inner Galaxy should not rotate like the bulge. 2) There is a causal connection between the formation of the halo and bulge, with the bulge being younger than the halo. 3) The metal–rich bulge globular clusters should not be older than the metal–poor halo globulars. It should be possible to test these predictions with the presently available technology.

I am happy to thank the best advisors in the Galaxy: Ed Olszewski, Jim Liebert, and Simon White, for years of help. This work was supported by the following grants: AST 91–19343, and AST 92–23967 to E. Olszewski, and AST 92–17961 to J. Liebert. I am also grateful for the help received from M. Irwin, J. Hill, M. Rich and A. McWilliam.

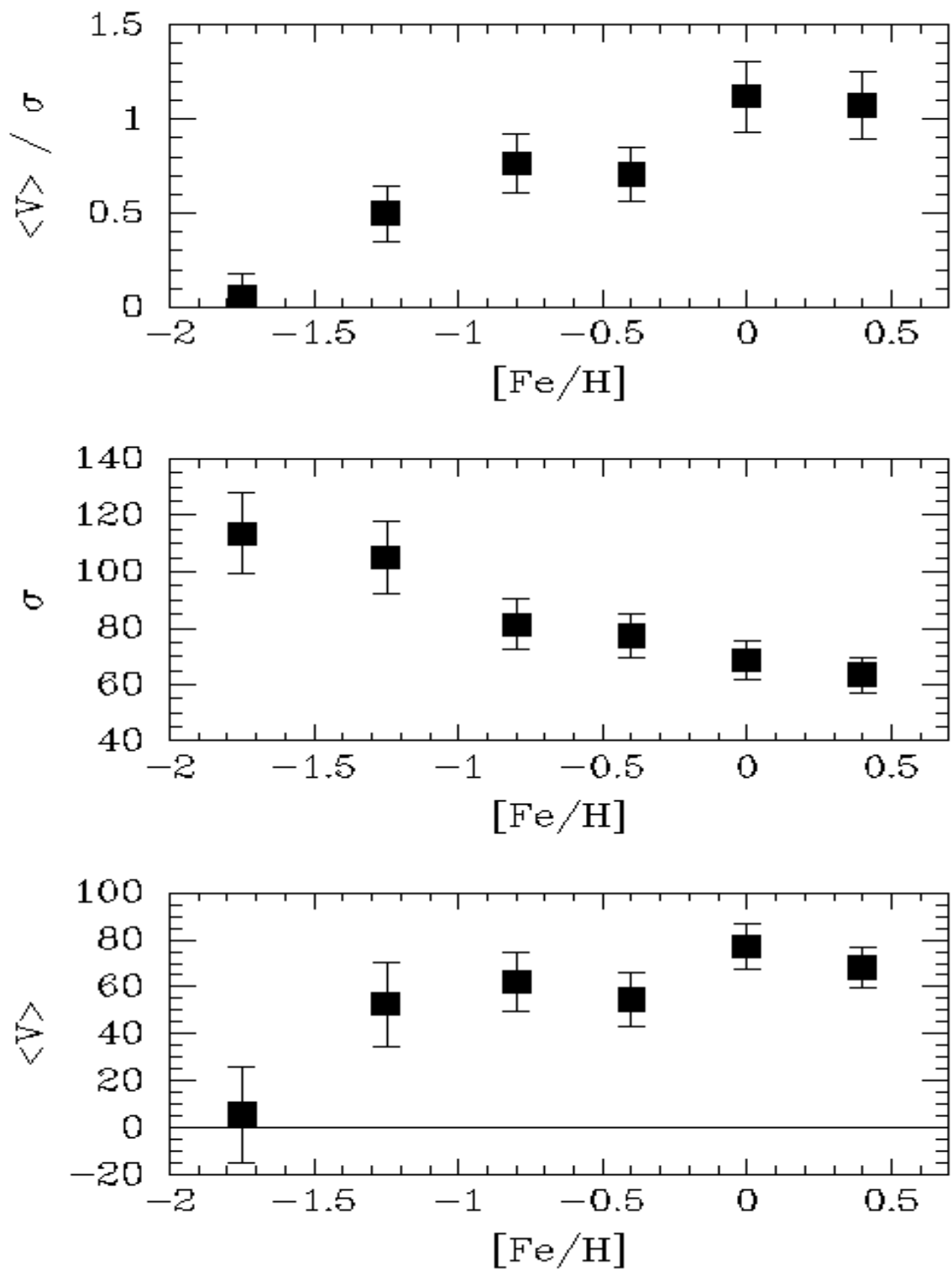

Fig. 1.— Mean Galactocentric velocities $V$, velocity dispersions $\sigma$, and local degree of rotational support $V/\sigma$ *vs* metallicity for K giants in the F588 bulge field, located at l,b = (8, 7), at 1.5 kpc projected distance from the Galactic center. Each bin includes 30 to 55 stars.

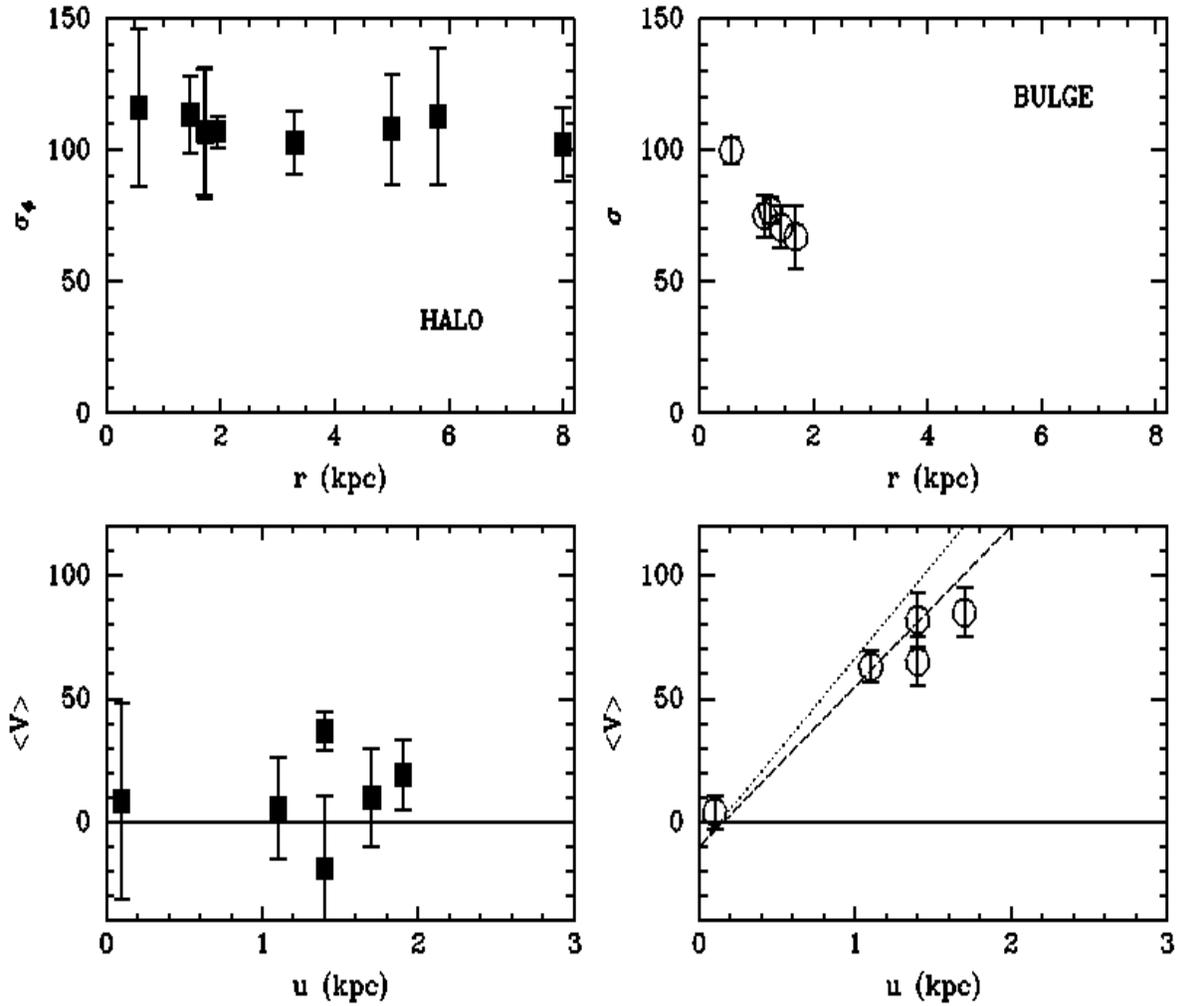

Fig. 2.— Run of the mean line of sight velocity $V$ *vs* Galactocentric distance projected in the plane u and velocity dispersions $\sigma$ *vs* Galactocentric distance for K giants with [Fe/H] $\leq -1.0$ (left panels), and with [Fe/H] $\geq -1.0$ (right panels) in different fields towards the Galactic bulge. The dashed line in the lower right panel shows the mean rotation of planetary nebulae from Kinman et al. (1990), and the dotted line shows the mean rotation of bulge Mira variables (Menzies 1990), and of bulge SiO masers (Nakada et al. 1993).

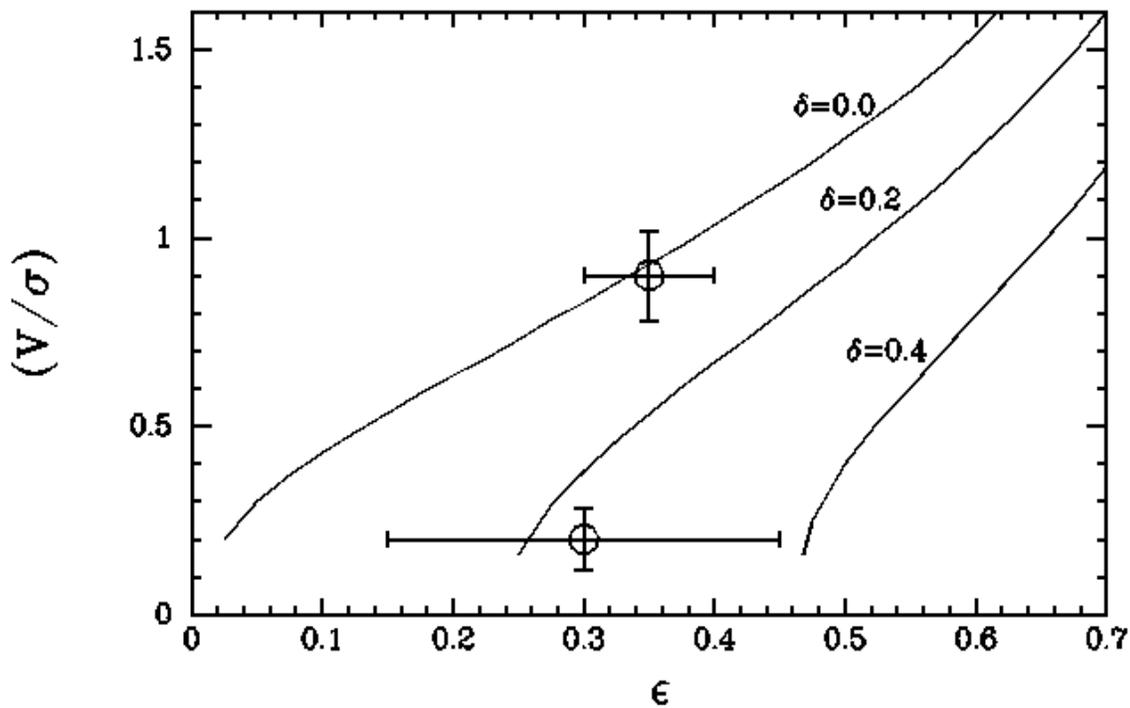

Fig. 3.— Relationship between rotation parameter $V/\sigma$ and ellipticity $\epsilon$ for coaxial oblate spheroids with different anisotropy $\delta$ (solid lines), reproduced from Binney & Tremaine (1987, their figure 4–5 in p. 217). We have plotted the location of the Galactic bulge (upper circle) and halo (lower circle). The bulge ellipticity is taken from Kent et al. (1991), and the halo ellipticity from the reviews by Bahcall (1986) and Freeman (1987). The 1 $\sigma$ error bars shown are conservative.